\documentclass[prl,twocolumn,showpacs,aps,floatfix,amsmath,amssymb,amsfonts,unsortedaddress]{revtex4-1}

\usepackage{graphicx}
\usepackage[normalem]{ulem}
\usepackage{color}
\usepackage{amsmath,amssymb}
\usepackage[T1]{fontenc}

\newcommand{\nhdp}{NH$_3^+$}

\newcommand{\nhtp}{NH$_2^+$}

\newcommand{\nhd}{NH$_3$}
\newcommand{\ec}{E$_\mathrm{coll}$}
\newcommand{\ecs}{E$_\mathrm{coll}$ }

\newcommand{\nhdps}{NH$_3^+$ }

\newcommand{\nhtps}{NH$_2^+$ }
\newcommand{\ndds}{ND$_3$ }
\newcommand{\nhds}{NH$_3$ }

\newcommand{\beq}{\begin{equation}}
\newcommand{\eeq}{\end{equation}}

\begin{document}
\title{Observation of orbiting resonances in He(3S1) + NH3 Penning ionization}

\author{Justin Jankunas$^1$, Krzysztof Jachymski$^2$, Micha\l\ Hapka$^3$, and Andreas Osterwalder$^1$\email{andreas.osterwalder@epfl.ch}}
\affiliation{$^1$Institute for Chemical Sciences and Engineering, Ecole Polytechnique F\'ed\'erale de Lausanne (EPFL), 1015 Lausanne, Switzerland\\
$^2$Faculty of Physics, University of Warsaw, Pasteura 5, 02-093 Warsaw, Poland\\
$^3$Faculty of Chemistry, University of Warsaw, Pasteura 1, 02-093 Warsaw, Poland}

\date{\today}
\begin{abstract}
A merged-beam study of the gas phase He($^3$S$_1$) + \nhd~Penning ionization reaction dynamics in the collision energy range 3.3 $\mu$eV $<$ \ec $<$ 10 meV is presented. 
In this energy range the reaction rate is governed by long-range attraction. 
Shape resonances are observed at collision energies of 1.8 meV and 7.3 meV and are assigned to $\ell$=15,16 and $\ell$=20,21 partial waves, respectively.
The experimental results, representing the first observation of shape resonances in a collision with a polyatomic molecule, are well reproduced by theoretical calculations with the short-range reaction probability $P_{sr}=0.035$.
\end{abstract}
\maketitle

\textit{Introduction.}
Chemical reactions of polyatomic molecules at low collision energies are interesting for a number of reasons. 
Studies of cold molecular collisions are an important source of information for interstellar chemistry.
From a fundamental point of view, an important motivation lies in the potential observation of quantum effects in molecular scattering, like quantum state-specific reactivity and scattering resonances~\cite{krems}.
Low energy collisions are characterised by few angular momentum partial waves. 
A prominent consequence of this is the possible observation of so-called partial-wave resonances, where a single partial wave can dominate the total reaction cross section~\cite{krems}.
In contrast, hundreds or even thousands of partial waves contribute in collisions at thermal energies, and the reaction can be treated statistically~\cite{levine}.
Resonances are among the clearest signatures of the quantum mechanical character of a molecular scattering event.
The first observation of orbiting resonances dates back to 1974 when the elastic scattering dynamics of H with Xe was studied~\cite{toennies:74}.
Low-energy scattering was not possible then, but this particular collision system, having a suitable interaction potential and low reduced mass, nevertheless enabled this important observation.
Until recently, that paper remained the only report on orbiting resonances in neutral reactions because for other collision systems it was not possible to reach the required temperature range.
Two recent developments, the merged beam technique and low-angle crossed beam scattering, have finally enabled scientists to venture into this regime and observe resonances in other reactions as well~\cite{Jankunas:2014JCP,Henson:2012kr,chefdeville:PRL}.

Studies of scattering resonances to date only targeted collisions involving atoms or diatomic molecules~\cite{Henson:2012kr,Narevicius:2014te,chefdeville:PRL,chefdeville:science}.
This paper reports the first observation of reactive scattering resonances in Penning Ionization (PI) of a polyatomic molecule with metastable helium. 
Specifically, the He($^3$S$_1$) +\nhds PI reaction is investigated in the collision energy (E$_{\mathrm{coll}}$) range 38 mK $<$ \ec $<$ 125 K. 
In contrast to atoms and diatomic molecules polyatomic molecules have a further reduced symmetry, and allow for a potentially more complex stereochemistry.
The enhanced complexity also potentially allows for a larger number of reaction channels, and it is not immediately clear how this affects a possible observation of resonances.

The first experiments on molecular PI were performed as early as 1970s \cite{Niehaus:1981vb}, but theoretical treatment of these electron transfer reactions has not been nearly as detailed as for the PI of atoms. 
PI at low \ecs has recently been studied using the merged-beam technique, where excited helium and neon atoms were collided with argon H$_2$\cite{Henson:2012kr, Narevicius:2014te}, ammonia~\cite{Bertsche:2014ub, Jankunas:2014JCP}, CH$_3$F~\cite{jankunas:jpc}, and CHF$_3$~\cite{jankunas:prep} at collision energies as low as 0.8 $\mu$eV (9 mK) where fewer than ten partial waves contribute to the collision. 
Partial wave resonances have been observed in He* + H$_2$/HD/D$_2$/Ar Penning ionization reactions, and were rather well reproduced theoretically~\cite{Narevicius:2014te}.

In the present reaction two channels are observed,
\begin{eqnarray}
\mathrm{He(^3S_1)} + \mathrm{NH}_3 &\rightarrow& \mathrm{NH}_3^+ + \mathrm{He} + e^- 	\label{eq:nodiss}\\
\mathrm{He(^3S_1)} + \mathrm{NH}_3 &\rightarrow& \mathrm{NH}_2^+ + \mathrm{H} + \mathrm{He} + e^-.\label{eq:diss}		
\end{eqnarray}
The internal energy of helium in the first excited $^3$S$_1$ state is sufficient to ionize \nhds through the ground $\tilde{\mathrm{X}}$ state of the ion or through the first excited state, the \nhdp($\tilde{\mathrm{A}}$) state, 4.8 eV above the ground state.
The $\tilde{\mathrm{X}}$ state is produced by removal of an electron from the lone pair and predominantly in the non-dissociating vibrational ground state~\cite{BenArfa:1999el}.
The $\tilde{\mathrm{A}}$ state, on the other hand, is formed by the removal of an electron from one of the N-H bonds and has a high probability to dissociate.
Based on this the authors of reference \onlinecite{BenArfa:1999el} suggest that the branching between the two channels is, in fact, a stereodynamic probe of this reaction.
The relative yield of the dissociative ionization (Eq. \ref{eq:diss}) in the related Ne($^3$P$_2$) + \nhds reaction has been observed to be constant over the entire range of collision energies from below 100 mK up to $\approx$250 K~\cite{Jankunas:2014JCP}.
At higher energies, the branching ratio for dissociative PI in the Ne*+\nhds reaction slightly increases \cite{BenArfa:1999el}.

\textit{Experimental setup.}
The  experimental apparatus and procedure is described in detail elsewhere~\cite{Jankunas:2014JCP}.
Briefly, electrostatic and magnetic hexapole guides are used to merge a beam of He($^3$S$_1$) atoms with a beam of \nhds molecules. 
Metastable helium atoms are produced in a commercial dielectric barrier discharge~\cite{luria:09}, mounted directly behind an Even-Lavie pulsed valve~\cite{even:valve}, operated with a backing pressure of 4 bars. 
Presumably two metastable helium states are produced, He($^1$S$_0$) and He($^3$S$_1$), but only the paramagnetic He($^3$S$_1$) species is guided, resulting in a very pure beam of atoms. 
The speed of He($^3$S$_1$) (denoted He* in the following) beam is controlled by cooling the valve in the range 140 K-- 200 K, resulting in continuously tuneable beam speeds in the range 1110 m/s -- 1420 m/s. 
Ammonia is prepared in a pulsed valve (general valve Series 9) with a backing pressure of 1.5 bar.
The speed of the ammonia beam is controlled by using pure \nhds or seeding it in Ne or Ar, resulting in beam velocities of approximately 1050 m/s, 820 m/s, or 620 m/s, respectively. 
In the electrostatic guide only low-field seeking (lfs) states are confined. 
Of the two states of the ammonia inversion doublet only the upper component is lfs while the lower component is high-field seeking and thus removed entirely from the beam.
The He* and \nhds beams are overlapped in space and time in the extraction region of a time-of-flight mass spectrometer (TOF-MS). 
\nhdps and \nhtps reaction products are detected by applying a 900 ns, -300 V pulse to the extraction plate of the TOF-MS when the desired velocities from the two expansions are inside. 
Product ions formed outside the TOF-MS are deflected and not detected.
The raw number of collected ions is normalized by the reactant densities that are measured separately.
The He* beam intensity is monitored by a multichannel plate detector mounted in the beam path, and the density of \nhd(J,K =1,1) is monitored by [2+1] resonance enhanced multiphoton ionization via the B(v$^\prime$=5)$\leftarrow$ X(v$^{\prime\prime}$=0) transition. 
The overall range in relative velocities covered at each data point, $\Delta v$, is $\approx$30 m/s~\cite{Jankunas:2014JCP}, limited by the pulse duration of about 60 $\mu$s of the general valve used for the supersonic expansion of \nhd. 
The conversion of relative velocity to collision energy means that $\Delta\mathrm{E_{coll}}$ is the smallest at the lowest energy and determines the minimum accessible collision energy. 
At the highest collision energies covered here $\Delta\mathrm{E_{coll}}$ amounts to several meV.

\textit{Measurement procedure.}
The energy dependent cross section $\sigma$(\ec) for the He* + \nhds Penning ionization reaction is obtained by converting the number of acquired ions according to 
\begin{equation}
\sigma \mathrm{v_{rel}}[\mathrm{NH}_3][\mathrm{He^*}]=\frac{\Delta \mathrm{NH}_3^+}{\Delta t}=k(\mathrm{E}_\mathrm{coll})[\mathrm{NH}_3][\mathrm{He^*}],\label{eq:rates}
\end{equation}
(and equivalently for \nhtp), where k(\ec) is the measured reaction rate coefficient, and v$_\mathrm{rel}$ is the relative velocity. 
Because the He* and \nhds beams co-propagate in a merged-beam experiment, the relative velocity is simply v$_\mathrm{rel}$ = v(He*) - v(\nhd). 
At the lowest collision energy the average relative velocity is zero, and the distribution of relative velocities is determined by the finite temporal width of the two gas pulses, yielding for the present experiment v$_\mathrm{rel} \approx$ 15 m/s~\cite{Jankunas:2014JCP}.

\begin{figure}[htpb]
 \includegraphics[width=\linewidth]{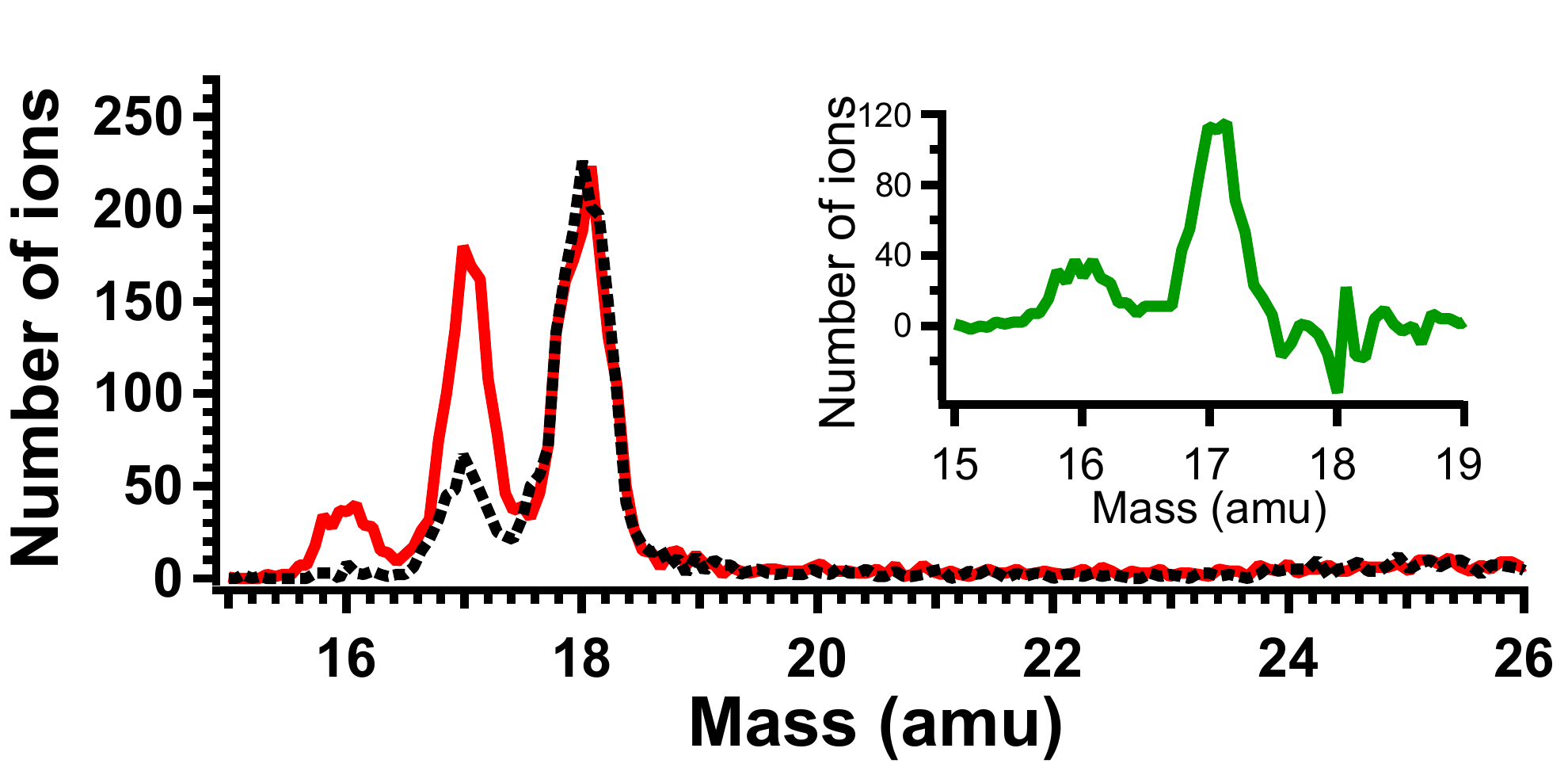}%
\caption{\label{fig:ms} Raw TOF mass spectrum of the He* + \nhds reaction. The solid red line shows the signal from temporally overlapped He* and \nhds beams. The dashed black trace is the background signal arising from the H$_2$O PI by He*, recorded when the \nhds beam is delayed by 3 ms with respect to the He* beam. The inset shows the reactive He* + \nhds signal after background subtraction.}
 \end{figure}
\begin{figure}
 \includegraphics[width=\linewidth]{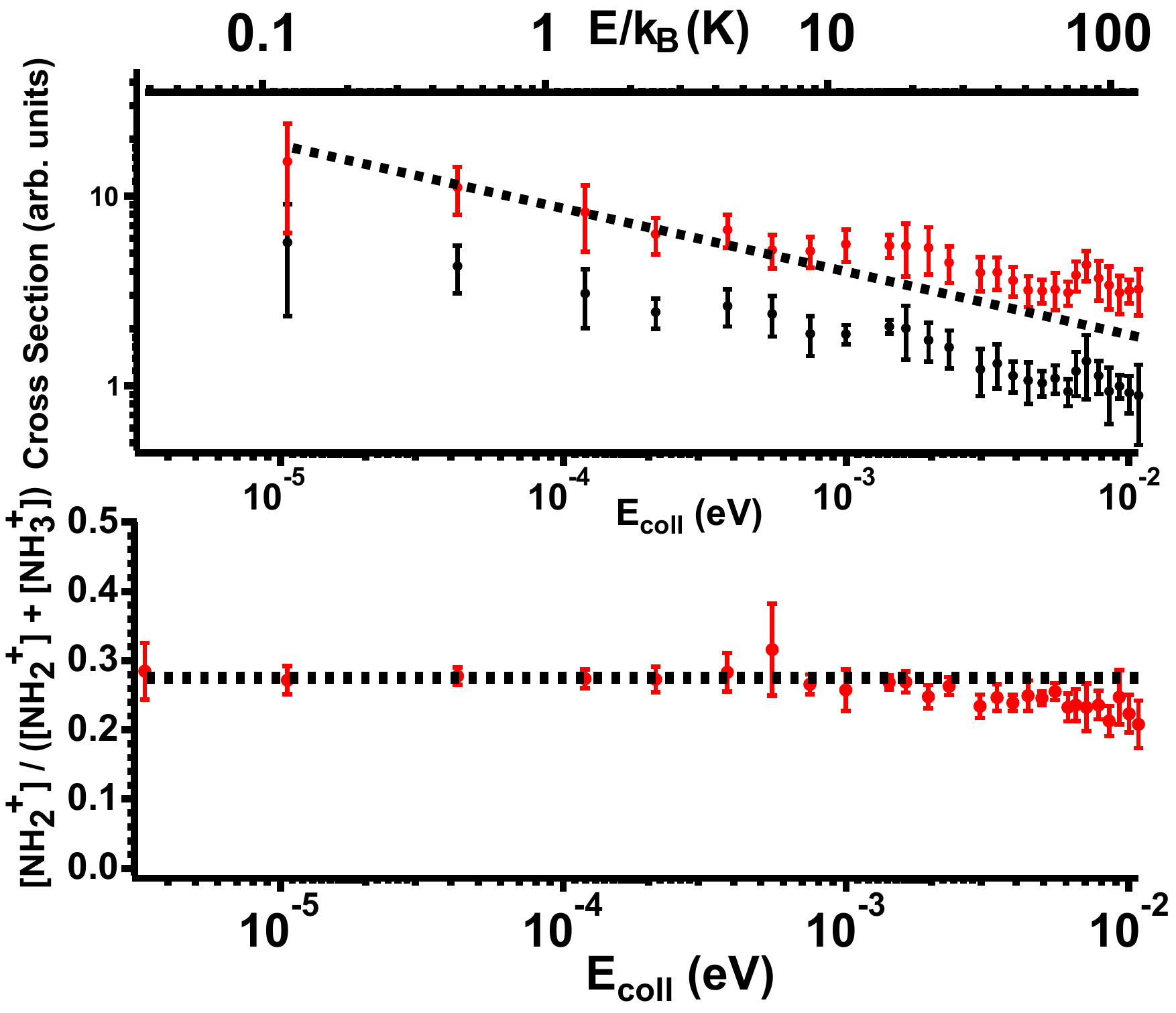}%
\caption{\label{fig:expt} Panel a: Experimental cross sections for the production of \nhdps (upper, red symbols) and \nhtps (lower, black symbols). The dashed line gives  the slope characteristic for a reaction dominated by long-range van der Waals interaction. Panel b: Branching ratio for the production of \nhtp, as defined in the text. The horizontal dashed line at $\Gamma=0.275$ serves to guide the eye. }
 \end{figure}
\begin{figure}
 \includegraphics[width=\linewidth]{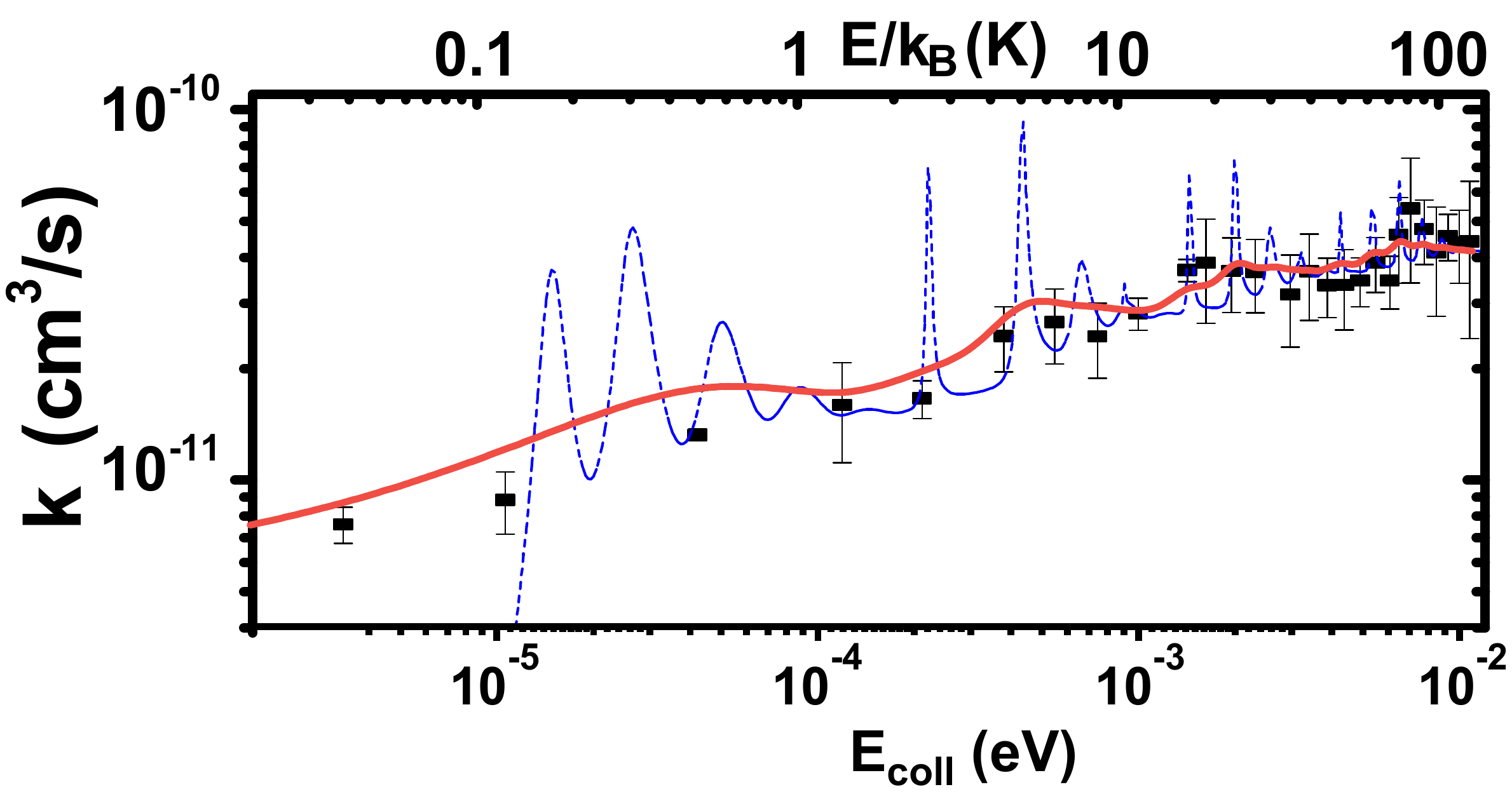}%
\caption{\label{fig:theo} Comparison of experimental (symbols) and theoretical reaction rate coefficients. The dashed line shows the raw theoretical data, the solid line after convolution with the experimental resolution.}
 \end{figure}

A typical mass spectrum is shown in Fig. \ref{fig:ms}. 
The main panel shows the ions collected under conditions where the two gas pulses overlap in time (solid red trace) and where the ammonia pulse is delayed by 3 ms relative to the He* pulse (black dashed line), respectively.
The dashed line shows two peaks, at masses 18 and 17, that correspond to (dissociative) Penning ionisation of background water.
The solid red trace also contains these ions but in addition, the desired products from the reaction with \nhds are present, namely \nhtps at 16 amu and \nhdps at mass 17.
The difference between these two traces, plotted in the inset, shows the background-corrected reactive signal of \nhtps and \nhdps product ions only. 
A complete excitation function is obtained by recording mass spectra like the one in Fig. \ref{fig:ms} for different beam velocities and integrating the ion signals for the respective masses.
Such a measurement is performed five times at each collision energy, and the error bars in the data below (Figures \ref{fig:expt}a and \ref{fig:theo}) are one standard deviation in the statistical fluctuations for these experiments.
Note that cross sections are given in arbitrary units because the determination of absolute reactant densities and beam overlap were not possible in this study.

\textit{Results.} The measured reaction cross section for the two He* + \nhds reactive channels is shown in Fig. \ref{fig:expt}a.
The branching ratio, defined as $\Gamma=\frac{[\mathrm{NH}_2^+]}{[\mathrm{NH}_2^+]+[\mathrm{NH}_3^+]}$, is displayed in Fig. \ref{fig:expt}b.     
The cross section of the He* + \nhds Penning ionization reaction is approximately proportional to $\mathrm{E_{coll}}^{-1/3}$ at the low energies measured here, as is indicated by the dashed black line in Fig. \ref{fig:expt}a.
A qualitatively similar behavior has been observed in the Ne* + \nhd~and Ne* + \ndds reactions~\cite{Jankunas:2014JCP}, and this is indeed expected when the interaction potential is assumed to follow a V(r)$\propto$ r$^{-6}$ dependence, and the reaction happens at short distance compared with the characteristic van der Waals length~\cite{levine}.
In this case, at low energies the cross section has the observed collision energy dependence $\sigma$(\ec) $\approx\mathrm{E_{coll}}^{-1/3}$~\cite{Niehaus:1981vb}.
When E$_\mathrm{coll}$ becomes comparable with the potential well depth, short range repulsive terms become important and the behavior of the rate constant should change.

Two maxima are observed in the reaction cross sections, at \ec=1.8 meV and 7.3 meV, that are due to tunnelling resonances through the centrifugal barrier with the formation of a transiently bound scattering state.
To understand the origin of these resonances theoretical calculations based on quantum defect theory (QDT) were performed.

The theoretical prediction of shape resonances based on {\it ab initio} dynamics is hindered by the difficulty in obtaining a full-dimensional potential energy surface (PES) for PI reaction of polyatomic systems (see, e.g. Ref.~\onlinecite{Siska:1993cf} and references therein). 
Nevertheless, modelling the experimental results is possible with the simplified approach applied to the description of the Ne*+\nhd~PI, i.e. QDT~\cite{Jankunas:2014JCP,Jachymski2013}.
 The necessary input for the QDT method is the knowledge of the long-range interaction. For the He*-NH$_3$ system the leading coefficients for dispersion $C_{6,{\rm disp}}^{00} =  265.45$ a.u. and induction $C_{6,{\rm ind}}^{00} = 105.44$  a.u. were obtained using symmetry-adapted perturbation theory (SAPT) \cite{Hapka:12} and suitable asymptotic expressions of the multipole expansion, respectively~\cite{Schmuttenmaer:91}.

Finally, in order to estimate the well depth of the He*-NH$_3$ complex the full {\it ab initio} potential energy curve for the collinear geometry of He*-NH$_3$ was studied  (see Ref.~\cite{Jankunas:2014JCP} for geometry description). The supermolecular coupled cluster singles and doubles (CCSD) calculations determined the global minimum to correspond to the He*-lone pair N arrangement, with a bond length of roughly 4.4 bohr and a well depth of 4114 cm$^{-1}$. 
A local minimum of 60.05 cm$^{-1}$ was found at roughly 9 bohr, corresponding to the He*-H$_3$N arrangement.
Those results are consistent with the findings of Ref.~\onlinecite{Pzuch:08} for the alkali atoms-NH$_3$ complexes.
Inversion doubling was neglected in the current calculations.
It must be kept in mind, though, that weak coupling between the two potential curves involving the (J,K)=(1,1) rotational level of \nhds might affect the reactivity at certain collision energies.

In QDT it is then noted that the reaction process takes place at distances much shorter than the ones characteristic for the long-range interaction, given by $R_6=\left(2\mu C_6/\hbar^2\right)^{1/4}\approx 46$ bohr.
In practice this allows to parametrize the wave function at short distances, either analytically for a pure van der Waals potential, or numerically for a more realistic one, and solve the resulting scattering problem. The key parameters of the model are the short-range reaction probability (P$_\mathrm{sr}$) and the short-range phase of the wave function (for a detailed description see Refs.~\onlinecite{Jankunas:2014JCP} and \onlinecite{Jachymski2013}).

In the present case, the finite depth of the attractive potential needs to be taken into account. A Lennard-Jones potential $V=-\frac{C_6}{r^6}+\frac{C_{12}}{r^{12}}$ is used for QDT calculations, with the $C_6$ coefficient taken from SAPT and $C_{12}$ adjusted to give the correct well depth. The positions and widths of the resonances can be varied in this simple model by manipulating P$_\mathrm{sr}$ and the scattering length. However, results obtained this way cannot reproduce the behavior of the data for the lowest collision energies, where the rate constant should approach a constant value, following the Wigner threshold laws~\cite{Sadeghpour2000}. A small potential barrier of $\approx 0.3$ cm$^{-1}$ which slightly suppresses the reaction at the lowest partial waves has to be added to obtain satisfactory agreement (see Figure~\ref{fig:theo}). 
Such barriers in general come from avoided crossings or anisotropic term couplings of the partial waves at short range.
A reason for such crossing could be the existence of the inversion doubling which has been neglected in the current study~\cite{Pzuch:09}.
It is possible to model this phenomenon by adding an additional channel to the QDT model, weakly coupled to the entrance.
The QDT simulations indeed reproduce a suppressed or enhanced reaction rate, depending on the coupling strength and phase shift in this additional channel, but for a detailed and accurate description of the current problem the model needs further extension.

Figure \ref{fig:theo} shows a comparison between the calculated reaction rate coefficients (lines) and the scaled experimental data (symbols).
The dashed blue line gives the raw results from QDT calculation with the potential described in the previous paragraphs.
The solid line gives the results convoluted with the experimental resolution.
The positions of the observed resonances are well reproduced.
The main contributions to the resonances come from the $\ell=15,16$ (at 1.8 meV) and $\ell=20,21$ (at 7.3 meV) partial waves, respectively. 
A third resonance, for $\ell=9$ at $\approx$ 0.5 meV, is predicted theoretically, and might also be present in the experimental data. 
The good agreement between experiment and theory indicates that He* + \nhds reaction is indeed dominated by van der Waals interactions. Fitting the short-range reaction probability to reproduce the observed features allows then to bring the experimental results to absolute scale. For the present reaction, P$_\mathrm{sr}\approx0.035$ was found. 
This value is very different from the ones measured in collisions of heavy polar molecules such as KRb+KRb~\cite{Ospelkaus2010}, for which P$_\mathrm{sr}$ is close to unity~\cite{Idziaszek2010}. 
The Langevin capture model predicts universal loss at short distances (P$_\mathrm{sr}$=1).
This means that there can not be any resonances, and the cross section can be predicted, for heavy systems with a large density of states, based on statistical models~\cite{Ruzic2013}. 
Deviations from this have been observed previously~\cite{Pzuch:09,Henson:2012kr,PSJ2012,Wang2013} but a complete model for such phenomena still needs to be developed.

Additional information on the dynamics of the process is obtained by inspecting the branching ratio of the \nhdp- and the \nhtp-channels, shown in figure \ref{fig:expt}b.
It remains almost constant from the lowest collision energies studied here, up to \ec$\approx$1 meV, above which it appears to slightly drop.
This is the energy range where the resonances are observed, but the branching ratio itself does not display any clear resonance structure.
Previous studies explained the observation of the two reaction channels by a stereodynamical effect: He* approaching along the N-lone pair axis predominantly produces \nhdp, while an approach along one of the N-H bonds produces predominantly \nhtp~\cite{BenArfa:1999el}.
The second case produces excited \nhdps in the $\tilde{A}$ state which dissociates.
The dominant channel produces stable ground state \nhdp.
The observed decrease in branching ratio corresponds to a slightly increased probability of \nhdp($\tilde{X}$) production.
This is interpreted as follows: at the position of the resonance He* and \nhds form a relatively long-lived complex.
Assuming the probability density in this state to be the highest in the He*-lone pair bound configuration, the complex would arrange in this way, irrespective of the initial angle of approach.
In the experiment this is observed as a global decrease of $\Gamma$(\nhtp), rather than local dips, which may simply be due to insufficient resolution.

\textit{Conclusions.}
Penning ionization of \nhds by He($^3$S$_1$) has been studied in a merged-beam experiment by measuring reaction cross sections at collision energies ranging from 3.3 $\mu$eV to 10 meV.
Two resonances were observed at \ec=1.8 meV and 7.3 meV.
QDT calculations, based on the long range interaction potential, reproduced the positions of the resonances well, thus allowing to assign them to specific angular momentum quantum numbers.
At collision energies $\approx$1 meV, the measured rate constant was found to differ from model predictions. A small potential barrier had to be added to the potential to obtain satisfactory agreement.
The branching ratio between \nhdps and \nhtps production was found constant up to a collision energy around 1 meV, above which the relative production of \nhtps drops slightly.
This is explained by assuming the resonance state to be localised mostly in a He*-lone pair bound configuration.
Increased reaction time then leads to increased probability for production of ground state \nhdp.

These results represent the first observation of shape resonances in a reaction that involves a polyatomic molecule, and they show that the merged beams technique is a powerful tool to study cold molecular reactions, also of relatively complex polyatomic molecules.
Despite this enhanced complexity, resonances were nevertheless observed.
The apparent independence of the position of these resonances from the reaction channel appears to indicate that the long range interaction potential in the entrance channel is almost isotropic, which effectively eliminates stereodynadmical effects.
 
The theoretical model basing on the long-range interaction was able to reproduce the experimental results. 
However, slight disrepancies at low energies show that a more sophisticated treatment, which would include details of the molecular structure and potential surfaces, would give significantly more information about the system and allow for better understanding of the reaction dynamics.

\textit{Acknowledgments.}
We thank Dr. Piotr \.{Z}uchowski for useful discussions.
Support from the Swiss National Science Foundation (grant number PP0022-119081) and EPFL is acknowledged. 
K.J. was supported by the Foundation for Polish Science International PhD Projects Programme co-financed by the EU European Regional Development Fund. 
M.H. was supported by the project ''Towards Advanced Functional Materials and Novel Devices: Joint UW and WUT International PhD Programme'', operated within the Foundation for Polish Science MPD Programme co-financed by the EU European Regional Development Fund and by the Polish Ministry of Science and Higher Education Grant No. N204 248440. %

%

\end{document}